\documentclass[prl,twocolumn,showpacs,amsmath,amssymb]{revtex4}
\usepackage{graphicx}
\usepackage{dcolumn}
\usepackage{bm}
\begin{document}
\title{
New Algorithm of the Finite Lattice Method 
for the High-temperature Expansion \\
of the Ising Model in Three Dimensions
}
\author{Hiroaki Arisue}
\affiliation{Osaka Prefectural College of Technology, 
        Saiwai-cho 26-12, Neyagawa, Osaka 572-8572, Japan}
\email{arisue@las.osaka-pct.ac.jp}

\author{Toshiaki Fujiwara}
\affiliation{Faculty of General Studies, Kitasato University, 
        Kitasato 1-15-1, Sagamihara, Kanagawa 228-8555, Japan}
\email{fujiwara@clas.kitasato-u.ac.jp}

\begin{abstract}
We propose a new algorithm of the finite lattice method to generate
the high-temperature series for the Ising model in three dimensions.
It enables us to extend the series for the free energy 
of the simple cubic lattice from the previous series of 26th order 
to 46th order in the inverse temperature. 
The obtained series give the estimate of the critical exponent 
for the specific heat in high precision.
\end{abstract}

\pacs{05.50.+q,02.30.Mv,75.10.Hk,75.40.Cx}

\maketitle
 The finite lattice method\cite{Enting1977,Creutz1991,Arisue1984} 
is a powerful tool to generate the high- and low- temperature series 
for the spin models in the infinite volume limit. 
It avoids the tedious work of counting all the diagrams 
in the graphical method and reduce the problem to the calculation 
of the partition function.
In two dimensions the total amount of the calculations 
for the finite lattice method 
increases exponentially with the maximum order $N$ of the series.
On the other hand in three dimensions 
the total amount of the calculations grows exponentially with $N^2$
and except for some cases\cite{Bhanot1994,Guttmann1994,Bhanot1992,Guttmann1993,
Arisue1994,Arisue1995,Arisue1993,Bhanot1993,Guttmann1994b} 
many of the expansion series have been calculated by the graphical method. 
Here we present a new algorithm 
of the finite lattice method for the high-temperature expansion 
in three dimensions in which the total amount of the calculation 
increase approximately exponentially with $N\log{N}$ and that enables us 
to generate the series to much higher orders than not only 
the standard algorithm of the finite lattice method 
but also the graphical method.

 In the finite lattice method to generate the high-temperature series
for the free energy in three dimensions 
we calculate the partition function 
$Z(l_x \times l_y \times l_z)$ for the finite size lattices with 
$2(l_x + l_y + l_z) \le N$ and define recursively\cite{Arisue1984}
\begin{eqnarray}
\phi(l_x \times l_y \times l_z)
&=&\log{[Z(l_x \times l_y \times l_z)]} \nonumber\\
&&\!\!\!\!\!\!\!\!\!\!\!\!\!\!\!\!\!\!\!\!\!\!
-\sum_{
\stackrel{\scriptstyle 
  l_x^{\prime}\le l_x,l_y^{\prime}\le l_y,l_z^{\prime}\le l_z,}
{\scriptstyle 
  l_x^{\prime} + l_y^{\prime} + l_z^{\prime}\ne l_x + l_y + l_z}
}
\phi(l_x^{\prime} \times l_y^{\prime} \times l_z^{\prime})\;. \label{eqn:phi}
\end{eqnarray}
Here we use the notation for the lattice size such that the 
$1\times 1\times 1$ lattice means the unit cube 
composed of $2\times 2\times 2$ sites.
The Boltzmann factor for each bond connecting the nearest neighbor sites 
$k$ and $k^{\prime}$
is expressed as
\begin{equation}
  \exp\left(\beta s_k s_{k^{\prime}}\right)
 = \cosh\left(\beta\right) \left(1 + t s_k s_{k^{\prime}} \right), 
    \label{eqn:Boltzmann}
\end{equation}
with $t=\tanh\left(\beta\right)$.
We define the bond configuration as the set of bonds to which
the factor $t s_k s_{k^{\prime}}$ in (\ref{eqn:Boltzmann}) 
is assigned while the factor $1$ is assigned to the other bonds
of the finite size lattice.
Non-vanishing contribution to the partition function comes only 
from the bond configuration in which the bonds form one or more closed loops.
Each of the closed loops is a polymer 
in the standard cluster expansion\cite{Muenster1981}.
Then the Taylor expansion of $\phi(l_x \times l_y \times l_z)$ 
with respect to $t$ 
includes the contribution from all the clusters of polymers 
in the standard cluster expansion 
that can be embedded into the lattice of $l_x \times l_y \times l_z$ 
but cannot be embedded into any of its rectangular sub-lattices 
$l_x^{\prime} \times l_y^{\prime} \times l_z^{\prime}$\cite{Arisue1984}.
The expansion series of the free energy density in the infinite
volume limit is given by
\begin{equation}
f=\sum_{2(l_x + l_y + l_z) \le N} \phi(l_x \times l_y \times l_z)
\label{eqn:free-energy}
\end{equation}
The expansion series of $\phi(l_x \times l_y \times l_z)$
starts from the term of $t^n$ with $n=2(l_x + l_y + l_z)$,
which comes from the cluster of a single polymer (one closed loop of bonds)
that have two intersections with any plane perpendicular 
to the lattice bonds. 
Thus it is enough to restrict the lattice sizes for the summation
in (\ref{eqn:free-energy})
to those that satisfy $2(l_x + l_y + l_z) \le N$
to obtain the series for $f$ to order $t^N$.

\begin{figure}[tb!]
\includegraphics{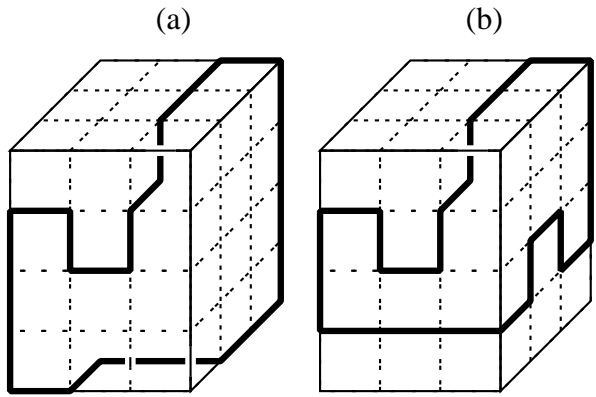}
\caption{\label{fig:config}
Examples of the bond configuration 
for $Z(l_x \times l_y \times l_z,\Delta N_z)$ 
with $l_x=3, l_y=3, l_z=4, \Delta N_z=2$.
Example (a) has $n_z{_1}=2$, $n_z{_4}=2$, $n_z{_3}=4$, and $n_z{_4}=2$
and it should be taken into account, while 
(b) has $n_z{_1}=0$, $n_z{_4}=4$, $n_z{_3}=4$, and $n_z{_4}=2$ 
and it should be neglected.
}
\end{figure}

In the standard algorithm of the finite lattice method 
the full partition function for the finite size lattice 
is calculated with all the bond configurations taken into account.
The point of the new algorithm is that,
in order to obtain the series to a given order, however, 
it is enough to consider only a restricted number of bond configurations.
Let us consider the anisotropic model of the simple cubic Ising model
with $\beta_i=J_i/k_BT$ and $t_i=\tanh{(\beta_i)}$ ($i=x,y,z$).
To obtain the series for 
$\phi(l_x \times l_y \times l_z)$ to order $N_z=2l_z+\Delta N_z$
in $t_z$ we introduce in the new algorithm
$\phi(l_x \times l_y \times l_z,\Delta N_z)$ 
defined recursively by 
\begin{eqnarray}
\phi(l_x \times l_y \times l_z,\Delta N_z)
&=&\log{[Z(l_x \times l_y \times l_z,\Delta N_z)]} \nonumber\\
&&\!\!\!\!\!\!\!\!\!\!\!\!\!\!\!\!\!\!\!\!\!\!
\!\!\!\!\!\!\!\!\!\!\!\!\!\!\!\!\!\!\!\!\!\!
-\sum_{\stackrel{\scriptstyle 
l_x^{\prime}\le l_x,l_y^{\prime}\le l_y,l_z^{\prime}\le l_z,}
       {\scriptstyle 
l_x^{\prime} + l_y^{\prime} + l_z^{\prime}\ne l_x + l_y + l_z}}
\phi(l_x^{\prime} \times l_y^{\prime} 
\times l_z^{\prime},\Delta N_z)\;. \label{eqn:Zd}
\end{eqnarray}
Here the partition function 
$Z(l_x \times l_y \times l_z,\Delta N_z)$ is calculated
only with the bond configurations taken into account 
that have orders $n_z{_i}$ in $t_z$ for the $i$-th layer perpendicular to the
z-axis ($i=1,2,\cdots,l_z$) satisfying 
\begin{equation}\sum_{i=1}^{l_z} \max(n_z{_i},2) \le 2l_z +\Delta N_z\;.
\end{equation} 
Examples of the bond configuration 
for $Z(l_x \times l_y \times l_z,\Delta N_z)$ 
that should be taken into account or should be neglected
can be seen in Fig.~\ref{fig:config}.

It is easy to prove that any bond configuration for the partition function
that has $\sum_{i=1}^{l_z} n_z{_i} \le 2l_z +\Delta N_z$
but $\sum_{i=1}^{l_z} \max(n_z{_i},2) > 2l_z +\Delta N_z$
does not contribute 
to the left hand side of (\ref{eqn:Zd}) in the order 
lower than or equal to $N_z = 2l_z+\Delta N_z$.
For such a bond configuration at least one of the $n_z{_i}$'s should be zero, 
so they are disconnected configuration(composed of more than one polymer)
or they can be embedded into a rectangular sub-lattice 
of the $l_x \times l_y \times l_z$ lattice 
and in either case they do not contribute 
to $\phi(l_x \times l_y \times l_z,\Delta N_z)$ 
in the order lower than or equal to $N_z = 2l_z +\Delta N_z$. 
They can contribute to $\phi(l_x \times l_y \times l_z,\Delta N_z)$ 
in higher order than $N_z = 2l_z +\Delta N_z$
by constituting the connected cluster of polymers
together with other configurations with $n^{\prime}{_z{_i}}\ge 2$ 
for the layer $i$ with $n_z{_i}=0\;$.
Then total order of the cluster of polymers has 
the order higher than $N_z = 2l_z +\Delta N_z$.

The contribution of the bond configuration with $\{n_z{_i}\}$
to the partition function of the finite size lattice 
can be calculated by the transfer matrix formalism as
\begin{equation}Z(\{n_z{_i}\})=V_{0,j_1}t_z^{n_z{_1}}V_{j_1,j_2}t_z^{n_z{_2}}
 \cdots t_z^{n_z{_{l_z}}} V_{j_{l_z},0}\,. \label{eqn:Zv}
\end{equation}
Here $V_{j_i,j_{i+1}}$ is the transfer matrix element 
with incoming $n_z{_i}$ spins and outgoing $n_z{_{\;i+1}}$ spins
and the summations over the spin locations $j_1,j_2,\cdots$ 
of the $n_z{_1}, n_z{_2}, \cdots$ spins, respectively, 
are assumed in the right hand side of (\ref{eqn:Zv}). 
This transfer matrix element itself is the partition function in two dimensions
with $n_z{_i} + n_z{_{\;i+1}}$ spins attached,
which can be calculated to any order in $t_x$ and $t_y$ efficiently 
by the site-by-site construction\cite{Enting1980,Bhanot1990}.
The amount of the calculation for each transfer matrix element
is proportional to the combinatorial factor 
$C(\ (l_x+1)(l_y+1)\ ,n_z{_i} + n_z{_{\;i+1}})$,  $2^{l_x}$ 
and $l_x l_y$, which 
are the number of the cases for 
attaching the $n_z{_{i}} + n_z{_{\;i+1}}$ spins to the $(l_x+1)(l_y+1)$ sites, 
the number of states in site-by-site construction 
for the partition function of the Ising model in two dimensions
and the number of the relevant bonds, respectively.

To obtain the series to order $N$ in the isotropic model,
we calculate the expansion series for each of 
the $\phi(l_x \times l_y \times l_z)$'s defined by (\ref{eqn:phi}) in the anisotropic model 
to order $t_x^{N_x}t_y^{N_y}t_z^{N_z}$ with $N_x+N_y+N_z=N$ 
using the new algorithm described above and set $t_x=t_y=t_z=t$ finally. 
When we use the new algorithm for each lattice size $l_x\times l_y\times l_z$ 
and  each of the orders $N_x, N_y, N_z$,
we can make the simultaneous exchange 
of the lattice axes and corresponding orders 
as $(l_x\times l_y\times l_z; N_x, N_y, N_z) 
\to (l_x\times l_z\times l_y; N_x, N_z, N_y)$ 
or $(l_x\times l_y\times l_z; N_x, N_y, N_z) 
\to (l_y\times l_z\times l_x; N_y, N_z, N_x)$ 
that have the chance to reduce the amount of the calculation 
for the transfer matrix elements.

\begin{figure}[tbp!]
\includegraphics{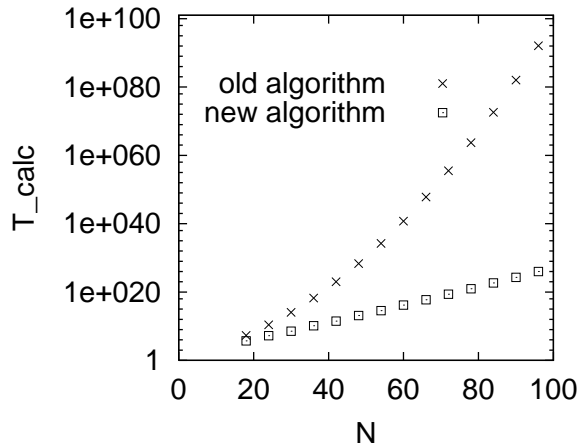}
\caption{\label{fig:calcsize}
Total amount of the calculation
to generate the series to order $N$
by the old algorithm and the new algorithm of the finite 
lattice method.
}
\end{figure}
We estimate the total amount of the calculation time $T_{\mathrm{calc}}(N)$
to generate the free energy series to order $N$ by listing up 
all the transfer matrices needed and summing up the estimated time to
calculate each of the matrices, which is plotted in Fig.~\ref{fig:calcsize} 
together with the estimated total calculation time
for the old algorithm of the finite lattice method.
The numerical estimation for the new algorithm agrees well with the
actually used calculation time for $N\le 46$.
We see that the calculation time for the old algorithm 
grows exponentially with $N^2$,
while that for the new algorithm can be fitted by 
$A\exp{(BN\log{N}+CN+D\log{N})}$ with $B\sim 0.15$, $C\sim -0.24$ and
$D\sim 3.6$. 
We can simply understand this $\exp( B N \log N +... )$ behavior
as follows.
The maximum size of the lattice to be taken into account 
is $l_x=l_y=l_z=N/6$,
for which $\Delta N_z=0$ and we have only to consider 
the bond configurations with $n_z{_i}=2$ for all $i=1,2,\cdots,l_z$. 
The largest amount of the calculation is to be paid
for the partition function of the lattice that has smaller size 
of $l_x\sim l_y\sim l_z\sim N/12$
for which the maximum of $n_z{_{i}} + n_z{_{\;i+1}}$ is about $N/6$ and 
the product of the above factors is approximately proportional 
to $\exp\left( BN\log{N}+CN+D\log{N} \right)$ with $B=1/6$ for large $N$. 

\begin{table}[tbp!]
\caption{\label{tab:series}
High-temperature expansion coefficients for the free energy density
of the simple cubic Ising model.
         }
\begin{center}
\begin{tabular}{rl}
$  n$ & $\qquad a_n$  \\
\hline
$  2$ & $\qquad  0                            $  \\ 
$  4$ & $\qquad  3                            $  \\ 
$  6$ & $\qquad  22                           $  \\ 
$  8$ & $\qquad  375/2                        $  \\ 
$ 10$ & $\qquad  1980                         $  \\ 
$ 12$ & $\qquad  24044                        $  \\ 
$ 14$ & $\qquad  319170                       $  \\ 
$ 16$ & $\qquad  18059031/4                   $  \\ 
$ 18$ & $\qquad  201010408/3                  $  \\ 
$ 20$ & $\qquad  5162283633/5                 $  \\ 
$ 22$ & $\qquad  16397040750                  $  \\ 
$ 24$ & $\qquad  266958797382                 $  \\ 
$ 26$ & $\qquad  4437596650548                $  \\ 
$ 28$ & $\qquad  525549581866326/7            $  \\ 
$ 30$ & $\qquad  6448284363491202/5           $  \\ 
$ 32$ & $\qquad  179577198475709847/8         $  \\ 
$ 34$ & $\qquad  395251648062268272           $  \\ 
$ 36$ & $\qquad  21093662188820520521/3       $  \\ 
$ 38$ & $\qquad  126225408651399082182        $  \\ 
$ 40$ & $\qquad  4569217533196761997785/2     $  \\ 
$ 42$ & $\qquad  291591287110968623857940/7   $  \\ 
$ 44$ & $\qquad  8410722262379235048686604/11 $  \\ 
$ 46$ & $\qquad  14120314204713719766888210   $  \\ 
\end{tabular}   
\end{center}
\end{table}
Using the new algorithm of the finite lattice method 
we have calculated the high-temperature series for the free energy
density of the simple cubic Ising model to order $N=46$.
The coefficients of the obtained series
\begin{equation}
f = 3 \cosh{(\beta)} + \sum_n a_n t^n ;\quad t = \tanh{(\beta)} 
\end{equation}
are listed in Table~\ref{tab:series}.
They agree with those given by Bhanot et al\cite{Bhanot1994} to order $N=24$
and by Guttmann and Enting\cite{Guttmann1994} to order $N=26$.
We have add ten new terms to these previous series.
To obtain the series of order $N=26$ in our new algorithm we needed only 
the computer memory of 1 MBytes and the CPU-time of 5 minutes 
in a standard PC and to order $N=46$ we used the total memory of 2 GBytes
and the CPU time of 25 hours in CP-PACS at Tsukuba University. 

\begin{figure}[tbp!]
\includegraphics{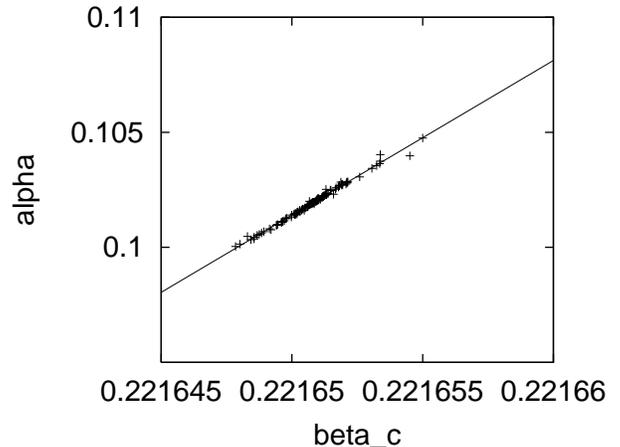}
\caption{\label{fig:alpha-t}
Critical exponent $\alpha$ versus the critical value $\beta_c$ 
for the inhomogeneous differential approximants
of the first order.
}
\end{figure}

Following is the result of the preliminary analysis of 
the series for the specific heat $C(t)=\sum_{n} c_n t^n$.
We plot in Fig.~\ref{fig:alpha-t} the critical exponent $\alpha$ 
versus the critical value $\beta_c$ 
for the first order inhomogeneous differential approximants
of the series of $N=40$--$46$. 
From the linear dependence of $\alpha$ on $\beta_c$ 
we find $\alpha=0.1045(1)$ or $\alpha=0.1077(2)$
at the value $\beta_c=0.22165459(10)$\cite{Blote1999} 
or $\beta_c=0.2216595(15)$\cite{Ito2000} obtained
in the recent Monte Carlo simulations.
The second order inhomogeneous differential approximants
give almost the same value.
\begin{figure}[tbp!]
\includegraphics{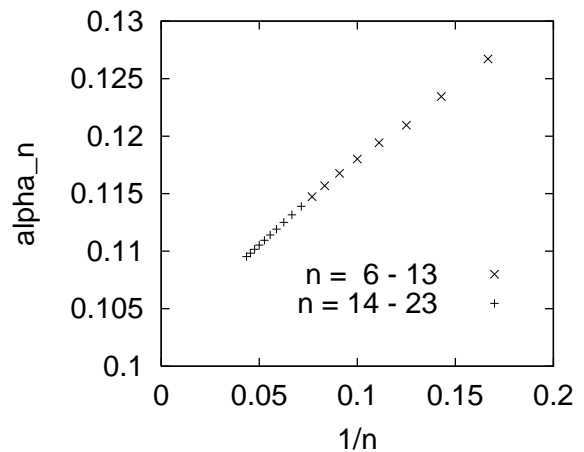}
\caption{\label{fig:alpha-n}
Sequence $\alpha_n$ plotted versus $1/n$.
}
\end{figure}
We also made the ratio analysis.
The sequence $\alpha_n=(t_c^2 c_{2n}/c_{2n-2}-1)n+1$,
which is expected to behave as $\alpha+b/n^{\Delta}+c/n+\cdots$ 
with the correction-to-scaling exponent $\Delta\sim 0.5$,
is plotted versus $1/n$ in Fig.~\ref{fig:alpha-n} for $\beta_c=0.22165459$.
The sequence for $n=14$--$23$ given by the new 10 coefficients
has a bit different slope from the sequence for $n\le 13$ given 
by the previously obtained series.
Three-parameter fitting of the newly obtained sequence
for $\beta_c=0.22165459$ gives 
$\alpha=0.1036(10)$, $b=-0.007(10)$ and $c=0.17(2)$ for $\Delta=0.5$. 
As for the case of $\beta_c=0.2216595$
it gives $\alpha=0.1082(14)$, $b=-0.033(10)$ and $c=0.21(2)$. 
These estimated values of $\alpha$ by the inhomogeneous differential 
approximation and by the ratio method are not inconsistent 
with each other.
We see, on the other hand, that the estimated values of $\alpha$ depends 
crucially on the value of $\beta_c$ and in order to determine $\alpha$ 
precisely in these biased method we need more precise value of $\beta_c$.
We also notice that the correction-to-scaling term is very small,
which was already pointed out in the analysis 
of the shorter series\cite{Bhanot1994,Guttmann1994}.

From the hyperscaling relation $\alpha=2-d\nu$ the high-temperature 
series for the second moment correlation length 
gives $\alpha=0.1088(39)$\cite{Butera2002} 
and $\alpha=0.1096(5)$\cite{Campostrini2002} 
and the $\epsilon$-expansion gives $\alpha=0.1088(39)$\cite{Guida1997}.
More direct estimation of $\alpha=0.110(2)$ was also given 
by the high-temperature series for the magnetic susceptibility of the
antiferromagnetic critical point\cite{Campostrini2002}.
We note that our estimated value of $\alpha$ using $\beta_c=0.22165459$ 
is not consistent with these values 
but the value using $\beta_c=0.2216595$ is rather closer
to these values. 

The basic idea presented here in the new algorithm of the finite lattice method 
can be applied to the high-temperature expansion of other quantities such as 
the magnetic susceptibility and the correlation length for the Ising model 
in three dimensions
and it can also be applied to the models with continuous spin variables 
such as the XY model in three dimensions.
Furthermore the idea can be used in the low-temperature expansion 
for the spin models in three dimensions.
We can expect that it will enable us to generate the expansion series
in much higher orders compared with the presently available series.

\paragraph*{Acknowledgments.---}
The work of H.A.\ is supported in part by Grant-in-Aid 
for Scientific Research (No.\ 12640382) 
from the Ministry of Education, Culture, Sports, Science and Technology.


\end{document}